\documentclass[twocolumn,showkeys,aps,prb,showpacs]{revtex4-1}
\usepackage{graphicx}
\usepackage[CJKbookmarks,dvipdfm,colorlinks,linkcolor=blue,citecolor=blue]{hyperref}

\begin{document}

\title{Potential thermoelectric material  $\mathrm{Cs_2[PdCl_4]I_2}$: a first-principles study}

\author{San-Dong Guo}
\affiliation{Department of Physics, School of Sciences, China University of Mining and
Technology, Xuzhou 221116, Jiangsu, China}
\begin{abstract}
The electronic structures and  thermoelectric properties of  $\mathrm{Cs_2[PdCl_4]I_2}$ are investigated by the  first-principles calculations and semiclassical Boltzmann transport theory. Both  electron and phonon transport are considered to attain the figure of merit $ZT$. A modified Becke and Johnson (mBJ) exchange potential, including spin-orbit coupling (SOC), is employed to investigate  electronic part of  $\mathrm{Cs_2[PdCl_4]I_2}$.
It is found that  SOC has obvious effect on valence bands, producing huge spin-orbital splitting, which leads to  remarkable detrimental effect on p-type  power factor. However,  SOC has a negligible influence on  conduction bands, so the n-type power
factor  hardly change. The temperature dependence of lattice thermal conductivity by assuming  an inverse
temperature dependence is attained  from reported ultralow lattice thermal conductivity of 0.31  $\mathrm{W m^{-1} K^{-1}}$ at room temperature. Calculating scattering time $\tau$ is challenging, but a hypothetical  $\tau$ can be adopted to estimate thermoelectric conversion efficiency.
The maximal figure of merit $ZT$ is up to about 0.70 and 0.60  with scattering time $\tau$=$10^{-14}$ s and   $\tau$=$10^{-15}$ s, respectively. These results  make us believe that  $\mathrm{Cs_2[PdCl_4]I_2}$ may be a potential thermoelectric material.
\end{abstract}


\maketitle

\section{Introduction}
Thermoelectric materials, which can  realize the direct  heat to electricity conversion and  make  essential contributions
to the crisis of energy, have attracted a great deal of attention\cite{s1,s2}.
A good thermoelectric material can be governed by the dimensionless  figure of merit $ZT=S^2\sigma T/(\kappa_e+\kappa_L)$, where S, $\sigma$, T, $\kappa_e$ and $\kappa_L$ are the Seebeck coefficient, electrical conductivity, absolute  temperature, the electronic and lattice thermal conductivities, respectively.  The high-performance thermoelectric materials should possess high $ZT$, which requires high power factor ($S^2\sigma$) and low thermal conductivity ($\kappa=\kappa_e+\kappa_L$).
The high lattice  thermal conductivity  is often a fatal disadvantage to gain high $ZT$ value like classic  half-Heusler thermoelectric materials\cite{s3,s4,s5,s6}.  The lattice  thermal conductivity can be reduced  by point defects  and nanostructuring\cite{bc1,bc2,bc3,bc4,bc5,bc6}.

 However, the ultralow thermal conductivity has been achieved  experimentally in SnSe single crystals, and an unprecedented $ZT$ of 2.6 at 923 K has been reported\cite{zhao}.
Theoretically,  Atsuto Seko et al. recently  discovered 221 materials with very low lattice  thermal conductivity,  $\mathrm{Cs_2[PdCl_4]I_2}$ of which  has  an electronic band gap of 0.88 eV calculated  within generalized
gradient approximation (GGA), and   ultralow lattice thermal conductivity of 0.31 $\mathrm{W m^{-1} K^{-1}}$ at 300K\cite{ltc1}.  The related electronic structure calculations of $\mathrm{Cs_2[PdCl_4]I_2}$ is very less. Recently, Li et al. studied  the electronic structures and thermoelectric properties of $\mathrm{Cs_2[PdCl_4]I_2}$, and predicted that $\mathrm{Cs_2[PdCl_4]I_2}$ is an indirect-band semiconductor with coexistence of several ionic and covalent bonds\cite{li}. However, the SOC is neglected, which has important effects on electronic structures for compound containing heavy element like I. The possible $ZT$  has not also been reported.
\begin{figure}
 \includegraphics[width=5cm]{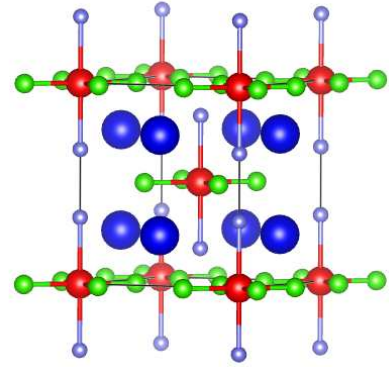}
 \caption{(Color online) The crystal structure of $\mathrm{Cs_2[PdCl_4]I_2}$. The largest  balls represent Cs atom, the following is Pd and Cl atoms, and the smallest  balls I atoms.}\label{st}
\end{figure}
\begin{table*}[!htb]
\centering \caption{ Peak $ZT$ for both n- and p-type  with $\tau$=$10^{-14}$ s and $\tau$=$10^{-15}$ s, and   the corresponding doping concentrations.  }\label{tab}
  \begin{tabular*}{0.96\textwidth}{@{\extracolsep{\fill}}ccccccccc}
  \hline\hline

                 &    & $\tau$=$10^{-14}$ s & & &   &  $\tau$=$10^{-15}$ s& &\\\hline
                   &n&&p&&n&&p&\\
  T (K) &($\mathrm{\times10^{19}cm^{-3}}$)&$ZT$&($\mathrm{\times10^{19}cm^{-3}}$)&$ZT$&($\mathrm{\times10^{19}cm^{-3}}$)&$ZT$&($\mathrm{\times10^{19}cm^{-3}}$)&$ZT$\\\hline\hline
600 &1.11&0.66&7.34&0.65&5.56&0.40&31.61&0.37\\
900 &1.54&0.71&6.07&0.72&4.26&0.55&25.64&0.52\\
1200 &6.60&0.63&19.07&0.69&7.48&0.58&27.85&0.60\\
1500&16.83&0.55&55.00&0.66&19.44&0.53&57.75&0.62\\\hline\hline
\end{tabular*}
\end{table*}

Here, we report on the thermoelectric properties of  $\mathrm{Cs_2[PdCl_4]I_2}$ from  a combination of  first-principles calculations and semiclassical Boltzmann transport theory.
The SOC  has been found to be very important for  power factor calculations in many thermoelectric materials\cite{so1,so2,so3,so4,so5,gsd3,gsd4,so6}, so the SOC  is considered  in our calculations of electronic part to attain reliable  power factor. As is well known, local density approximation (LDA) and GGA underestimate semiconductor energy gaps, and an improved mBJ exchange potential is used to investigate electronic structures of $\mathrm{Cs_2[PdCl_4]I_2}$.
It is found that SOC has  a noteworthy reduced influence on p-type power factor, which can be understood by considering SOC effects on valence bands. The ultralow lattice thermal conductivity is a key factor to attain high $ZT$, and the corresponding  room temperature lattice thermal conductivity of $\mathrm{Cs_2[PdCl_4]I_2}$ is only 0.31  $\mathrm{W m^{-1} K^{-1}}$\cite{ltc1}, which can be used to attain temperature dependence of lattice thermal conductivity by assuming  an inverse temperature dependence. Finally, the dimensionless thermoelectric figure of merit $ZT$   can be estimated by assuming  $\tau$=$10^{-14}$ s or $\tau$=$10^{-15}$ s, and the $ZT$ can be up to about  0.70  or 0.60 at about 1000 K  by the optimized doping.

The rest of the paper is organized as follows. In the next section, we shall
describe computational details. In the third section, we shall present the electronic structures and  thermoelectric properties of  $\mathrm{Cs_2[PdCl_4]I_2}$. Finally, we shall give our discussions and conclusion in the fourth
section.
\begin{figure}
  \includegraphics[width=8.0cm]{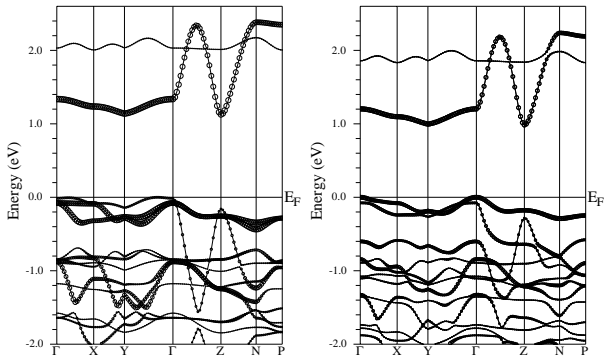}
  \caption{The energy band structures of $\mathrm{Cs_2[PdCl_4]I_2}$  using mBJ (Left) and mBJ+SOC (Right), and the dot diameter is proportional to the I atom weight at each k-point.}\label{band}
\end{figure}

\begin{figure*}
  \includegraphics[width=15cm]{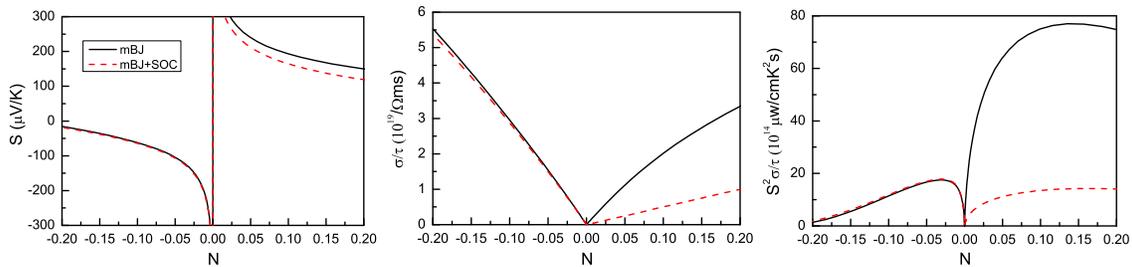}
  \caption{(Color online)  At room temperature (300 K),  transport coefficients of $\mathrm{Cs_2[PdCl_4]I_2}$ as a function of doping level (N):  Seebeck coefficient S, electrical conductivity with respect to scattering time  $\mathrm{\sigma/\tau}$  and   power factor with respect to scattering time $\mathrm{S^2\sigma/\tau}$   using  mBJ (Black solid lines) and mBJ+SOC (Red dash lines). The doping level (N) implies  electrons (minus value) or holes (positive value) per unit cell.}\label{t1}
\end{figure*}
\begin{figure*}
  \includegraphics[width=15cm]{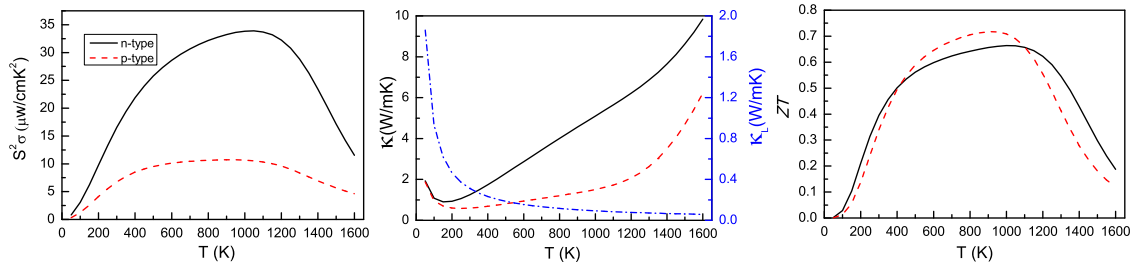}
  \caption{(Color online) The  power factor  $\mathrm{S^2\sigma}$,  thermal conductivities  (total thermal conductivity  $\mathrm{\kappa}$ and lattice  thermal conductivity  $\mathrm{\kappa_L}$) and $ZT$ as a function of temperature with the doping concentration of $\mathrm{5\times10^{19}cm^{-3}}$ for n-type and p-type, and the scattering time $\mathrm{\tau}$  is 1 $\times$ $10^{-14}$ s.   The doping concentration equals  $\mathrm{3.3518\times10^{21}cm^{-3}}$ $\times$ doping level.}\label{s1}
\end{figure*}

\begin{figure}
  \includegraphics[width=7cm]{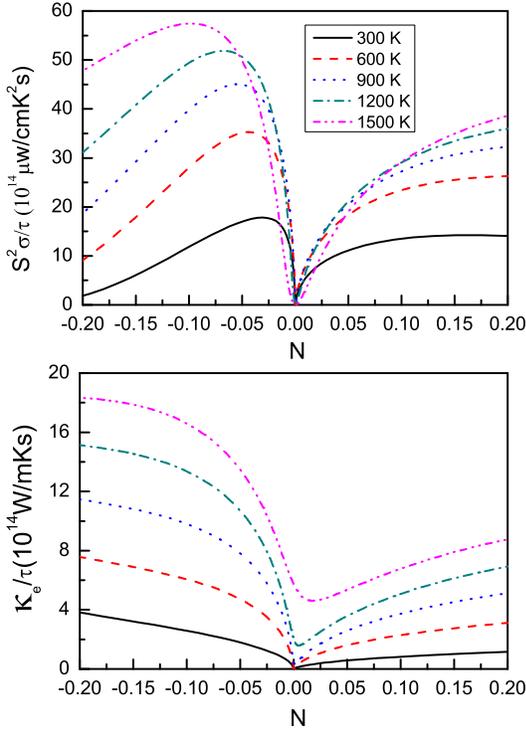}
  \caption{(Color online) The power factor with respect to scattering time  $\mathrm{S^2\sigma/\tau}$ and  electronic thermal conductivity with respect to scattering time $\mathrm{\kappa_e/\tau}$  of  $\mathrm{Cs_2[PdCl_4]I_2}$ as a function of doping level (N) with temperature  being 300, 600, 900, 1200 and 1500 (unit: K) using mBJ+SOC. }\label{t2}
\end{figure}
\begin{figure}
  \includegraphics[width=7cm]{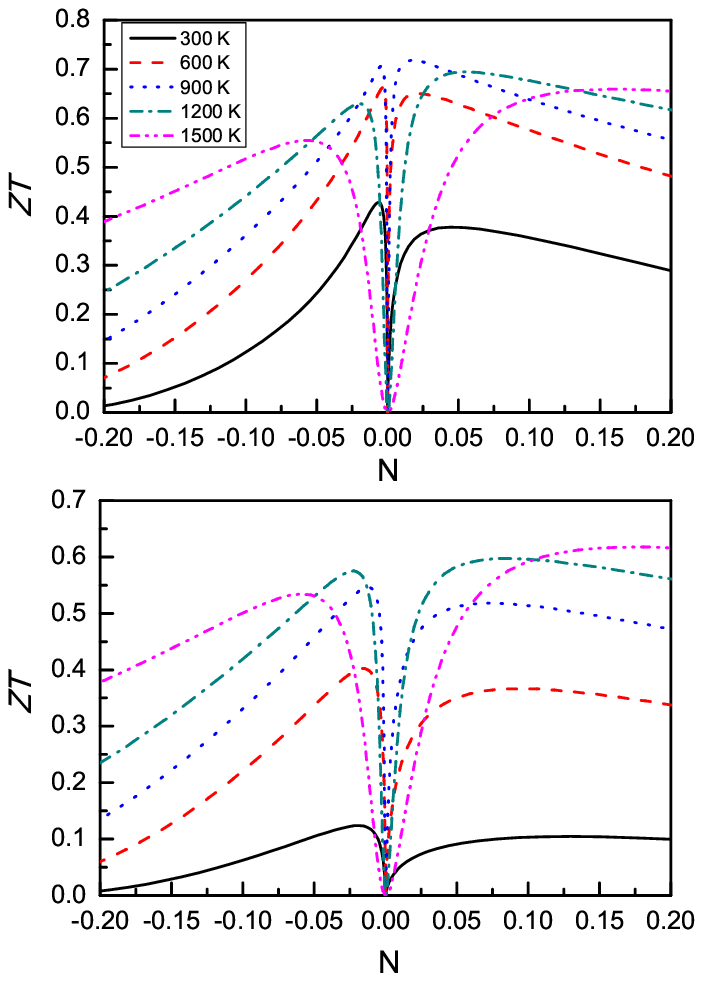}
  \caption{(Color online) The $ZT$ of  $\mathrm{Cs_2[PdCl_4]I_2}$ as a function of doping level with temperature  being 300, 600, 900, 1200 and 1500 (unit: K), and the scattering time $\mathrm{\tau}$  is 1 $\times$ $10^{-14}$ s (Top) and 1 $\times$ $10^{-15}$ s (Bottom). }\label{t4}
\end{figure}
\section{Computational detail}
The electronic structures of $\mathrm{Cs_2[PdCl_4]I_2}$ are performed using
a full-potential linearized augmented-plane-waves method
within the density functional theory (DFT) \cite{1}, as implemented in
the WIEN2k  package\cite{2}.  We employ Tran and Blaha's mBJ
 exchange potential plus LDA
correlation potential  for the
exchange-correlation potential \cite{4}, which has been known to produce
more accurate band gaps than LDA and GGA.
 The SOC was included self-consistently \cite{10,11,12,so} due to containing heavy elements, which leads to band splitting, and produces important effects on power factor. We use 5000 k-points in the
first Brillouin zone for the self-consistent calculation,  make harmonic expansion up to $\mathrm{l_{max} =10}$ in each of the atomic spheres, and set $\mathrm{R_{mt}*k_{max} = 8}$. The self-consistent calculations are
considered to be converged when the integration of the absolute
charge-density difference between the input and output electron
density is less than $0.0001|e|$ per formula unit, where $e$ is
the electron charge. Transport calculations, including Seebeck coefficient, electrical conductivity and electronic
 thermal conductivity,
are performed through solving Boltzmann
transport equations within the constant
scattering time approximation (CSTA) as implemented in
BoltzTrap\cite{b}, and reliable results  have
been obtained for several materials\cite{b1,b2,b3}. To
obtain accurate transport coefficients, we use 50000 k-points (36$\times$36$\times$36 k-point mesh) in the
first Brillouin zone for the energy band calculation.

\section{MAIN CALCULATED RESULTS AND ANALYSIS}
$\mathrm{Cs_2[PdCl_4]I_2}$ belongs to tetragonal phase crystal
structure with space group $I4/mmm$, which is shown in \autoref{st}.
The experimental lattice parameters (a=b=8.15 $\mathrm{\AA}$, c=8.99 $\mathrm{\AA}$)\cite{icsd} are used, and the atomic positions are optimized within GGA.  A improved mBJ exchange potential is used to investigate the electronic structures of $\mathrm{Cs_2[PdCl_4]I_2}$, which has been proved to be very effective to accurately calculate gaps of all kinds of semiconductors\cite{4,41,42}. Due to containing heavy element I, it is very crucial for electronic structure calculations to consider SOC.
For comparison, the projected energy band structures with both mBJ and mBJ+SOC are shown in \autoref{band}. Calculated results show that $\mathrm{Cs_2[PdCl_4]I_2}$ is a indirect gap semiconductor using both mBJ and mBJ+SOC, and the corresponding gap value is 1.13 eV and  0.99 eV. The mBJ gap value is larger than the reported GGA value of 0.88 eV\cite{ltc1}. A noticeable feature of conduction bands is that conduction band minimum  (CBM)  at Z point and conduction band subminimum at Y point is almost degenerate, namely band convergence\cite{s1,gsd3},  which is  benefit for power factor.
The projected band structures show that the first three valence bands and first conduction band have obvious I atom character. It is found that SOC has remarkable influence on the  valence bands, while has a  negligible effect on conduction bands. The SOC leads to giant  spin-orbital splitting of the first three valence bands, which produces remarkable effect on power factor of $\mathrm{Cs_2[PdCl_4]I_2}$.

The  semi-classic transport coefficients are performed  using CSTA Boltzmann theory.
The doping effects are simulated  by shifting the Fermi level within the framework of the rigid band approach, which has been proved to be reasonable in the low doping level\cite{tt9,tt10,tt11}.
 At room temperature, the  Seebeck coefficient S,  electrical conductivity with respect to scattering time  $\mathrm{\sigma/\tau}$ and  power factor with respect to scattering time $\mathrm{S^2\sigma/\tau}$  as  a function of doping level  using mBJ and mBJ+SOC are shown  in \autoref{t1}.
 The n-type doping (negative doping levels) with the negative Seebeck coefficient can be imitated by shifting Fermi level into the conduction bands.  When the Fermi level moves into valence bands, the p-type doping (positive doping levels) with the positive Seebeck coefficient can be achieved. When the SOC is considered, both S and  $\mathrm{\sigma/\tau}$ are smaller than ones without SOC in p-type doping, while they are nearly the same for n-type.  Therefore,  $\mathrm{S^2\sigma/\tau}$ using mBJ+SOC becomes
 very small compared to one with mBJ in p-type doping, but it changes almost nothing in n-type doping.
At the absence of SOC, p-type best power factor is much larger than n-type one. However, including SOC, n-type best power factor
is larger than p-type one. Similar SOC influence on best power factor can be observed in $\mathrm{Mg_2Sn}$\cite{gsd3}.

The SOC effect on S can be explained  by the following formula\cite{eq1}:
\begin{equation}\label{f1}
  S=\frac{\pi^2}{3}(\frac{k_B^2T}{e})[\frac{1}{n}\frac{dn(E)}{dE}+\frac{1}{\mu}\frac{d\mu(E)}{dE}]_{E=E_f}
\end{equation}
where $k_B$, $e$,  $n(E)$, $\mu(E)$, $E_f$ are the Boltzmann constant,   carrier charge,
 carrier density at the energy E,   mobility, and   Fermi energy, respectively.
In the valence bands (p-type doping), it is found that $\frac{dn(E)}{dE}$ with mBJ+SOC becomes small with respect to one with mBJ due to remarkable spin-orbit splitting, leading to  SOC-reduced S. In conduction bands (n-type doping), $\frac{dn(E)}{dE}$ is
almost the same between mBJ and mBJ+SOC, which leads to nearly the same S. Due to  $\mathrm{\sigma}$ being proportional to $n$,
the $n$ with SOC is smaller than one without SOC for p-type, leading to reduced $\mathrm{\sigma}$, and they are almost the same for n-type, producing nearly the same $\mathrm{\sigma}$.

To attain $ZT$ needs  scattering time $\tau$, and calculating $\tau$ is challenging from the first-principle calculations. Here, we assume that $\tau$ equals 1 $\times$ $10^{-14}$ s. Another key parameter is lattice thermal conductivity $\kappa_L$, and the room temperature lattice thermal conductivity has been reported for 0.31  $\mathrm{W m^{-1} K^{-1}}$\cite{ltc1}. An inverse
temperature dependence of the lattice thermal conductivity can be found in a large
number of thermoelectric materials\cite{gsd3,gsd4,lc1,lc2}. Here, we  simply assume that $\kappa_L$ is proportional to $1/T$ also for $\mathrm{Cs_2[PdCl_4]I_2}$.  The  power factor  $\mathrm{S^2\sigma}$,  total  thermal conductivity $\kappa$,  lattice thermal conductivity $\kappa_L$  and $ZT$ as a function of temperature with the doping concentration of $\mathrm{5\times10^{19}cm^{-3}}$ for n-type and p-type are plotted in \autoref{s1}. The $\mathrm{S^2\sigma}$  in both n- and p-type doping firstly increases, and then decreases, when the temperature increases. The total thermal conductivity $\mathrm{\kappa}$ is dominated by the lattice thermal conductivity $\mathrm{\kappa_L}$ in the low temperature, but the electronic thermal conductivity $\mathrm{\kappa_e}$ becomes very larger than lattice thermal conductivity $\mathrm{\kappa_L}$  in  high temperature region.
So, the related temperature of minimum thermal conductivity is very low due to the
ultralow lattice thermal conductivity. Both p-type  $\mathrm{S^2\sigma}$  and  $\kappa$ are smaller than n-type ones in considered temperature range. The $ZT$ has the same trend with $\mathrm{S^2\sigma}$ with the increasing temperature. At about 1000 K, the peak $ZT$ of 0.66 is attained in n-type doping. The maximal p-type $ZT$ is 0.72 at about 950 K. The p-type $ZT$ is larger than n-type one from about 400 K  to 1100 K.

In order to further understand the thermoelectric properties of  $\mathrm{Cs_2[PdCl_4]I_2}$, the power factor with respect to scattering time  $\mathrm{S^2\sigma/\tau}$ and  electronic thermal conductivity with respect to scattering time $\mathrm{\kappa_e/\tau}$  of  $\mathrm{Cs_2[PdCl_4]I_2}$ as a function of doping level (N) with temperature  being 300, 600, 900, 1200 and 1500 using mBJ+SOC are shown in \autoref{t2}. In the considered doping and temperature range,  the $\mathrm{S^2\sigma/\tau}$ firstly increases, and then decreases  in low doping level with increasing temperature, while it monotonically increases in high doping level. However, $\mathrm{\kappa_e/\tau}$ always  monotonically increases, when the temperature increases.
 The corresponding thermoelectric figure of merit $ZT$  with hypothetical $\tau$=$10^{-14}$ and  $\tau$=$10^{-15}$ s are plotted in \autoref{t4}.  As the temperature increases, the $ZT$ has similar trend with $\mathrm{S^2\sigma/\tau}$.
 The peak $ZT$ and corresponding  doping concentration at different temperature for both n- and p-type are listed in \autoref{tab}. It is found that n-type peak $ZT$ has lower  doping concentration than p-type one.  The  doping concentration of peak $ZT$ with  $\tau$=$10^{-15}$ s  is larger than one with $\tau$=$10^{-14}$ s. Calculated results also show that n- and p-type have almost the same  maximal $ZT$, and the maximal $ZT$ with $\tau$=$10^{-14}$ s  and $\tau$=$10^{-15}$ s is up to about  0.70 and 0.60, respectively.

\section{Discussions and Conclusion}
The bands of  $\mathrm{Cs_2[PdCl_4]I_2}$ near the Fermi level are dominated by heavy element I, which produces
a giant SOC in the valence bands.
The SOC  can lift the band degeneracy  of valence bands by  spin-orbit splitting, especially for the first three valence bands.
These SOC effects on valence bands  lead to remarkable reduced  influence  on p-type Seebeck coefficient and electrical conductivity, and  further give rise to  detrimental influence on power factor. The similar SOC-induced reduced effects on power factor have been observed  in many thermoelectric materials\cite{so1,so2,gsd3,gsd4}. Therefore, including SOC is very  necessary  for electronic part of  thermoelectric properties of $\mathrm{Cs_2[PdCl_4]I_2}$.

In summary,  an appropriate exchange-correlation potential mBJ+LDA is chosen to investigate  electronic structures and electronic part of thermoelectric properties  of $\mathrm{Cs_2[PdCl_4]I_2}$, and the SOC is also considered due to containing heavy element I.
The strength of  SOC effects on  valence bands  is very huge, especially for the first three valence bands, which leads to obvious reduced effects on p-type power factor.
The lattice thermal conductivity $\kappa_L$ of $\mathrm{Cs_2[PdCl_4]I_2}$  is assumed to be  proportional to $1/T$, and  the  $\kappa_L$ as a function of temperature is attained from the reported room temperature $\kappa_L$. Although the scattering time $\tau$ is unknown, a hypothetical $\tau$ can be employed to estimate possible figure of merit $ZT$, which
 is up to about 0.70 with $\tau$=$10^{-14}$, and about 0.60 with $\tau$=$10^{-15}$. Experimentally, it is possible to attain higher $ZT$ than theoretical values.
 The present work provides a foundation for further experimental studies.

\begin{acknowledgments}
This work is  supported by the Fundamental Research Funds for the Central Universities (2015XKMS073). We are grateful to the Advanced Analysis and Computation Center of CUMT for the award of CPU hours to accomplish this work.
\end{acknowledgments}


\begin{references}

\bibitem{s1} Y. Pei, X. Shi, A. LaLonde, H. Wang, L. Chen and G. J. Snyder, Nature \textbf{473}, 66 (2011).

\bibitem{s2}L. E. Bell,  Science \textbf{321}, 1457 (2008).


\bibitem{s3}J. R. Sootsman, Angew. Chem. \textbf{48}, 8616 (2009).

\bibitem{s4}M. Schwall and B. Balke, Appl. Phys. Lett. \textbf{98}, 042106 (2011).


\bibitem{s5} S.  Chen and Z. F. Ren,   Mater. Today   \textbf{16},   387 (2013).

\bibitem{s6}J. Yang, H. M. Li, T.  Wu, W. Q. Zhang, L. D.  Chen  and J. H. Yang, Adv. Funct. Mater.  \textbf{18}, 2880 (2008).


\bibitem{bc1}S. Bhattacharya, M. J. Skove, M. Russell, T. M. Tritt, Y. Xia, V. Ponnambalam, S. J. Poon and N. Thadhani
Phys. Rev. B \textbf{77}, 184203  (2008).
%Effect of boundary scattering on the thermal conductivity of TiNiSn-based half-Heusler alloy

\bibitem{bc2}V. Ponnambalam, P. N. Alboni, J. Edwards, T. M. Tritt, S. R. Culp and S. J. Poon, J. Appl. Phys. \textbf{103}, 063716 (2008).

%Thermoelectric properties of pp-type half-Heusler alloys Zr1?xTixCoSnySb1?yZr1?xTixCoSnySb1?y (0.0<x<0.50.0<x<0.5; y=0.15y=0.15 and 0.3)

\bibitem{bc3}P. F. Qiu, X. Y. Huang, X. H. Chen and L. D. Chen, J. Appl. Phys. \textbf{106}, 103703 (2009).

%Enhanced thermoelectric performance by the combination of alloying and doping in TiCoSb-based half-Heusler compounds

\bibitem{bc4} N. Shutoh and S. Sakurada, J. Alloy. Compd. \textbf{389},  204 (2005).

%Thermoelectric properties of the TiX(Zr0.5Hf0.5)1 ? XNiSn half-Heusler compounds

\bibitem{bc5}S. Bhattacharya, T. M. Tritt, Y. Xia, V. Ponnambalam, S. J. Poon and N. Thadhani, Appl. Phys. Lett. \textbf{81}, 43 (2002).
%Grain structure effects on the lattice thermal conductivity of Ti-based half-Heusler alloys


\bibitem{bc6}X. Yan, G. Joshi, et al. Nano Lett.  \textbf{11}, 556 (2011).
%Enhanced Thermoelectric Figure of Merit of p-Type Half-Heuslers

\bibitem{zhao}L. D.  Zhao,	S. H. Lo,	Y. S. Zhang  et al, Nature \textbf{508}, 373 (2014).

\bibitem{ltc1}A. Seko, A. Togo, H. Hayashi, K. Tsuda, L. Chaput and I. Tanaka, Phys. Rev. Lett. \textbf{115}, 205901 (2015).


\bibitem{li}W. F. Li and G. Yang, EPL \textbf{113}, 57007 (2016).

\bibitem{so1}K. Kutorasinski, B. Wiendlocha, J. Tobola and S. Kaprzyk,  Phys. Rev. B \textbf{89}, 115205 (2014).

\bibitem{so2}S. D. Guo, J. Alloy. Compd. \textbf{663}, 128 (2016).

\bibitem{so3} P. Larson, S. D. Mahanti, and M. G. Kanatzidis, Phys. Rev. B
\textbf{61}, 8162 (2000).

\bibitem{so4} T. J. Scheidemantel, C. Ambrosch-Draxl, T. Thonhauser, J. V.
Badding, and J. O. Sofo, Phys. Rev. B \textbf{68}, 125210 (2003).

\bibitem{so5}S. J. Youn and A. J. Freeman, Phys. Rev. B \textbf{63}, 085112 (2001).



\bibitem{gsd3}S. D. Guo and J. L. Wang, RSC Adv. \textbf{6}, 31272 (2016).

\bibitem{gsd4}S. D. Guo, RSC Adv. \textbf{6}, 47953 (2016).

\bibitem{so6}N. Singh and U. Schwingenschl$\ddot{o}$gl, Phys. Status Solidi RRL \textbf{08},
805 (2014).




%%%%%%%%%%%%%%%%%%%%%%%%%%%%%%%%%%%%%%%%%%%%%%%%%%%%%%%%%%%%%%%%%%%%
\bibitem{1}P. Hohenberg and W. Kohn, Phys. Rev. \textbf{136},
B864 (1964); W. Kohn and L. J. Sham, Phys. Rev. \textbf{140},
A1133 (1965).

\bibitem{2}P. Blaha, K. Schwarz, G. K. H. Madsen, D. Kvasnicka
 and J. Luitz, WIEN2k, an Augmented Plane Wave
+ Local Orbitals Program for Calculating Crystal Properties
(Karlheinz Schwarz Technische Universit\"at Wien, Austria) 2001,
ISBN 3-9501031-1-2
\bibitem{4}F. Tran and P. Blaha, Phys. Rev. Lett. \textbf{102},
226401 (2009).

%\bibitem{pbe}J. P. Perdew, K. Burke and M. Ernzerhof, Phys. Rev. Lett. \textbf{77}, 3865 (1996).

\bibitem{10}A. H. MacDonald, W. E. Pickett and D. D. Koelling, J. Phys. C \textbf{13}, 2675 (1980).

\bibitem{11}D. J. Singh and L. Nordstrom, Plane Waves, Pseudopotentials and the LAPW
Method, 2nd Edition (Springer, New York, 2006).

\bibitem{12}J. Kunes, P. Novak, R. Schmid, P. Blaha and
K. Schwarz, Phys. Rev. B \textbf{64}, 153102 (2001).

\bibitem{so}D. D. Koelling, B. N. Harmon, J. Phys. C: Solid State Phys.  \textbf{10}, 3107 (1977).



\bibitem{b}G. K. H. Madsen and D. J. Singh, Comput. Phys. Commun. \textbf{175}, 67
(2006).

\bibitem{b1}B. L. Huang and M. Kaviany, Phys. Rev. B \textbf{77}, 125209 (2008).

\bibitem{b2}L. Q. Xu, Y. P. Zheng and J. C. Zheng, Phys. Rev. B \textbf{82}, 195102 (2010).

\bibitem{b3}J. J. Pulikkotil, D. J. Singh, S. Auluck, M. Saravanan, D. K. Misra, A. Dhar and R. C. Budhani,
Phys. Rev. B \textbf{86}, 155204 (2012).


%%%%%%%%%%%%%%%%%%%%%%%%%%%%%%%%%%%%%%%%%%%%%%%%%%%%%%%%%%%%%%%%%%%%%%%%%%%%%%%%%%%%%%%%%%%%%
\bibitem{icsd} The experimental crystal structure is attained from the Inorganic Crystal Structure Database (ICSD).

\bibitem{41} S. D. Guo, J. Phys. D: Appl. Phys. \textbf{48}, 445004 (2015).

\bibitem{42}D. Koller, F.  Tran and P.  Blaha, Phys. Rev. B \textbf{83}, 195134  (2011).


\bibitem{tt9} T. J. Scheidemantel, C. Ambrosch-Draxl, T. Thonhauser, J. V. Badding and
J. O. Sofo, Phys. Rev. B \textbf{68}, 125210   (2003).


\bibitem{tt10} G. K. H. Madsen, J. Am. Chem. Soc. \textbf{128}, 12140 (2006).

\bibitem{tt11} X. Gao, K. Uehara, D. Klug, S. Patchkovskii, J. Tse and T. Tritt, Phys.
Rev. B \textbf{72}, 125202 (2005).

\bibitem{eq1}J. P. Heremans et al., Science \textbf{321}, 554 (2008).


\bibitem{lc1}D. Parker and D. J. Singh,  Phys. Rev. B \textbf{82}, 035204 (2010).

\bibitem{lc2}J. J. Pulikkotil  et al.,  Phys. Rev. B \textbf{86}, 155204 (2012).

\end{references}
\end{document}